%% file: main_v4.tex
\documentclass[10pt,pra,aps,twocolumn,superscriptaddress]{revtex4-1}
\usepackage[dvipdfmx]{graphicx}
\usepackage{amssymb,amsmath,amsthm,booktabs,mathtools}
\usepackage{bbm}
\usepackage{bm}
\usepackage{color}
\usepackage{ulem}
\usepackage{tikz}
\usepackage{hyperref}
\usepackage{enumitem}
\hypersetup{
    pdftitle={main},    
    pdfauthor={Scopa-Ruggiero-Calabrese-Dubail},
    colorlinks=true,
    citecolor=blue,
    linkcolor=red,
    urlcolor=blue
  }
\usepackage{times}
\usepackage{microtype}

\definecolor{ggray}{rgb}{0.9, 0.9, 0.9}
\def\tit#1{}
\usepackage{color}

\newcommand{\be}{\begin{equation}}
\newcommand{\ee}{\end{equation}}
\usepackage{ulem}

\renewcommand{\emph}{\textit}
\usepackage{amsmath}

\begin{document}
\newcommand{\titleinfo}{One-particle density matrix and momentum distribution of the out-of-equilibrium 1D Tonks-Girardeau gas:  Analytical results at large $N$
}
\title{\titleinfo}

\author{S. Scopa}
\affiliation
{SISSA and INFN, Via Bonomea 265, 34136 Trieste, Italy}
\author{P. Ruggiero}
\affiliation
{King’s College London, Strand WC2R 2LS, UK}
\author{P. Calabrese}
\affiliation
{SISSA and INFN, Via Bonomea 265, 34136 Trieste, Italy}
\affiliation
{International Centre for Theoretical Physics (ICTP), Strada Costiera 11, 34151 Trieste, Italy}
\author{J. Dubail}
\affiliation
{Laboratoire de Physique et Chimie Th\'eoriques, CNRS, UMR 7019, Universit\'e de Lorraine, 54506 Vandoeuvre-les-Nancy, France
}

\begin{abstract}
	In one-dimensional (1D) quantum gases, the momentum distribution (MD) of the atoms is a standard experimental observable, routinely measured in various experimental setups. The MD is sensitive to correlations, and it is notoriously hard to compute theoretically for large numbers of atoms $N$, which often prevents direct comparison with experimental data. Here we report  significant progress on this problem for the 1D Tonks-Girardeau (TG) gas in the asymptotic limit of large $N$, at zero temperature and driven out of equilibrium by a quench of the confining potential. We find  an exact analytical formula for the one-particle density matrix $\langle \hat{\Psi}^\dagger(x) \hat{\Psi}(x') \rangle$ of the out-of-equilibrium TG gas in the $N \rightarrow \infty$ limit, valid on distances $|x-x'| $ much larger than the interparticle distance. By comparing with time-dependent Bose-Fermi mapping numerics, we demonstrate that our analytical formula can be used to compute the out-of-equilibrium MD with great accuracy for a wide range of momenta (except in the tails of the distribution at very large momenta). For a quench from a double-well potential to a single harmonic well,which mimics a  `quantum Newton cradle' setup, our method predicts the periodic formation of peculiar, multiply peaked, momentum distributions.
\end{abstract}

\maketitle

\section{Introduction}
In the field of ultracold quantum gases, the momentum distribution (MD) of atoms has been a key experimental observable since the early studies of three-dimensional Bose-Einstein condensates~\cite{davis1995bose,anderson1995observation,stenger1999bragg}. It can be measured by Bragg spectroscopy~\cite{stenger1999bragg,richard2003momentum,fabbri2011momentum}, time of flight \cite{bourdel2003measurement,regal2005momentum,kinoshita2006quantum,stewart2010verification,wilson2020observation,malvania2021generalized} or focusing~\cite{shvarchuck2002bose,davis2012yang,jacqmin2012momentum,fang2016momentum}. 
The MD is the Fourier transform of the one-particle density matrix (1PDM) $\langle \hat{\Psi}^\dagger( {\bf x} ) \hat{\Psi}({\bf x}') \rangle$,
\be
n({\bf p}) =  \int d^d {\bf x} \int d^d {\bf x}' e^{\frac{i}{\hbar} {\bf p} ({\bf x}-{\bf x}')}  
\langle \hat{\Psi}^\dagger({\bf x}) \hat{\Psi}({\bf x}') \rangle,
\label{MD1}
\ee
where the second-quantized operators $\hat{\Psi}^\dagger({\bf x})/\hat{\Psi}({\bf x})$ create or destroy one atom at position ${\bf x}$. 
The MD is sensitive to non-local correlations in the gas~\cite{richard2003momentum,gerbier2003momentum,fabbri2011momentum}, especially in one-dimensional (1D) clouds where the effects of fluctuations and correlations are enhanced and destroy long-range order~\cite{schultz1963note,lenard1964momentum,petrov2000regimes,mora2003extension,kheruntsyan2003pair,rigol2004universal,cazalilla2004bosonizing}. Correlations in one dimension manifest themselves in various ways in the MD, for instance as a singularity at zero temperature, $n(p) \propto |p|^{1/2K -1}$ when $p \rightarrow 0$ ~\cite{cazalilla2004bosonizing,cazalilla2011one} (the dimensionless constant $K$ is the Luttinger parameter which parametrizes the interaction strength~\cite{giamarchi2003quantum}). The MD is also a key observable out of equilibrium, and in one dimension it often differs completely from its equilibrium counterpart. For instance, in the quantum Newton's cradle (QNC)~\cite{kinoshita2006quantum}, the MD of bosons colliding in a quasi-harmonic trap evades equilibration, even after thousands of collisions. Also, when a gas of interacting bosons is allowed to expand under a 1D geometry, the MD evolves non-trivially and, after long expansion times, becomes identical to the distribution of rapidities (or asymptotic momenta) of the initial state~\cite{sutherland1998exact,jukic2008free,jukic2009momentum,campbell2015sudden,caux2019hydrodynamics,bouchoule2022generalized}, a phenomenon known as `{dynamical fermionization}'~\cite{rigol2005fermionization,rigol2005free,minguzzi2005exact,dupays2022tailoring}, which allows to measure rapidity distributions~\cite{wilson2020observation,malvania2021generalized,Kuan2022}.\\[.1cm]
\indent
The theoretical calculation of the MD of strongly correlated atomic gases is a notoriously hard problem. In one dimension, where many experiments are described by the Lieb-Liniger model of bosons with contact repulsion~\cite{lieb1963exact,yang1969thermodynamics} or by one of its fermionic/multi component extensions~\cite{gaudin1967systeme,gaudin2014bethe,guan2013fermi}, it is generally not possible to access the dynamics of the MD by direct numerical simulations for large numbers of atoms $N$ and long times. Quantum Monte-Carlo calculations of the MD~\cite{jacqmin2012momentum,xu2015universal,fang2016momentum} are restricted to equilibrium, while time-dependent density matrix renormalization group simulations~\cite{peotta2014quantum,ruggiero2020quantum} or form factors resummations~\cite{caux2009correlation,caux2007one,konik2007numerical,panfil2014finite,caux2019hydrodynamics} are always restricted to short times and small numbers of particles. This has prevented direct modeling of experimental data for the MD in out-of-equilibrium setups~\cite{kinoshita2006quantum,malvania2021generalized}.\\[.1cm]
\indent
The situation is more favorable in the {\it Tonks-Girardeau} (TG) limit of hard-core bosons (infinite contact repulsion), where an efficient numerical evaluation of the 1PDM, and therefore also of the MD, can be obtained exploiting a time-dependent version of Bose-Fermi mapping (BFM)~\cite{girardeau1960relationship,girardeau2000breakdown,minguzzi2022strongly,rigol2004universal,rigol2005fermionization,pezer2007momentum,atas2017exact}. 
\\[.1cm]
\indent
On the analytical side, the search for exact solutions for the 1PDM and the MD of the TG gas is a long-standing challenge, see Refs.~\cite{schultz1963note,lenard1964momentum,vaidya1979one,jimbo1980density}  and e.g.~Sec.~III.A of Ref.~\cite{cazalilla2011one} for a review of this problem. Pioneering works from the 1960s and 1970s~\cite{schultz1963note,lenard1964momentum,vaidya1979one} focused on the ground state of the homogeneous TG gas and determined the asymptotic behavior of the 1PDM $ \langle  \Psi^\dagger(x) \Psi(x') \rangle  \propto |x-x'|^{-\frac{1}{2}}$ for $|x-x'| \gg L/N$ where $L$ is the system's length --a result that is also obtained in Luttinger liquid theory~\cite{cazalilla2004bosonizing,cazalilla2011one}. The case of a trapped gas with inhomogeneous density profile is harder, and the first analytical results for the ground state in a harmonic trap were obtained by Forrester-Frankel-Garoni-Witte only in the 2000s~\cite{forrester2003finite,papenbrock2003ground,gangardt2004universal}, while the case of an arbitrary trapping potential was cracked in 2017~\cite{brun2017one,colcelli2018universal} thanks to a new `{inhomogeneous Luttinger liquid}' approach~\cite{dubail2017conformal,dubail2017emergence,brun2018inhomogeneous,scopa2020one,gluza2022breaking,moosavi2022exact,tajik2022experimental}. Out of equilibrium, analytical results for the 1PDM and the MD have so far been limited to the dynamics in a harmonic trap with a time-dependent frequency~\cite{minguzzi2005exact,scopa2018exact,ruggiero2019conformal}; in that special case the 1PDM is related to the ground state one by a dynamical symmetry~\cite{eliezer1976note}. A crucial open problem in this area is the derivation of analytical results for more general quench dynamics.\\[.1cm]
This is precisely the purpose of this paper. Below we report an analytical formula for the out-of-equilibrium 1PDM of the TG gas, valid at large $N$ and for arbitrary trapping potentials, which is then used to evaluate the dynamics of the MD after the quench. Our analytical formula captures the behavior of the MD for a wide range of momenta (except in the tails of the distribution at very large momenta), complementing known results from Tan's contact physics \cite{minguzzi2002high,olshanii2003short,rigol2004universal,vignolo2013universal,decamp2016high,yao2018tan,bouchoule2021breakdown}.
\section{Model and quench protocol}
The Hamiltonian of the TG gas (with particle mass $=1$) in a trapping potential $V(x)$ is
\begin{equation}
	\label{eq:ham}
	\hat{H}=\int dx ~\hat{\Psi}^\dagger(x) \big(  - \frac{\hbar^2 \partial_x^2}{2} + V(x) + \frac{g}{2} \hat{\Psi}^\dagger(x) \hat{\Psi}(x)    \big)  \hat{\Psi}(x) ,
\end{equation}
with $[ \hat{\Psi}(x),  \hat{\Psi}^\dagger(y) ] = \delta(x-y)$, and repulsion coupling $g \rightarrow+\infty$. In that limit, two bosons cannot be at the same position and thus display fermionic-like properties. Under the Jordan-Wigner mapping to fermionic operators $\hat{\Psi}_{\rm F}^\dagger (x)=\exp \left( i \pi \int_{y<x} \hat{\Psi}^\dagger (y) \hat{\Psi}(y) dy \right) \Psi^\dagger(x)$, the Hamiltonian (\ref{eq:ham}) becomes quadratic
$\hat{H}=\int dx~\hat{\Psi}_{\rm F}^\dagger(x) \left(-\frac{1}{2}\hbar^2 \partial_x^2 + V(x) \right)\hat{\Psi}_{\rm F}(x) $ and  local quantities (such as density and current profiles) behave as non-interacting fermions~\cite{girardeau1960relationship}. The same does not apply to the 1PDM. In particular, the 1PDM of hard-core bosons is non-local in the fermionic basis
\be
\langle \hat{\Psi}^\dagger(x)   \hat{\Psi}(x')  \rangle =   \langle \hat{\Psi}^\dagger_{\rm F} (x)  e^{i\pi \int_{x}^{x'}dy~\hat{\Psi}^\dagger_{\rm F}(y) \hat{\Psi}_{\rm F}(y)}  \hat{\Psi}_{\rm F} (x')  \rangle
\ee
 and thus differs from the one of non-interacting fermions, $\langle \hat\Psi^\dagger_F(x)\hat\Psi_F(x')\rangle$.\\

In the following, we focus on the case where the TG gas is prepared in the ground state in an arbitrary trapping potential $V(x)=V_0(x)$. At time $t=0$, the dynamics is generated by suddenly changing the trapping potential from $V_0(x)$ to an arbitrary $V_1(x)$, a situation routinely realized in modern cold atom experiments~\cite{schemmer2019generalized,wilson2020observation,wilson2020observation,malvania2021generalized}.
\section{Hydrodynamic approach}
Our strategy for the calculation of the time-dependent bosonic 1PDM can be summarized as follows:
\begin{itemize}
\setlength\itemsep{1pt}
\item[\emph{i)}]~In this section, we establish the hydrodynamic evolution of the gas in terms of its Wigner function, related to non-interacting fermions. 
\item[\emph{ii)}]~In Sec.~\ref{sec:ana-res}, we shall include long-range Gaussian quantum fluctuations on top of the hydrodynamic background to determine $\langle\hat\Psi^\dagger(x)\hat\Psi(x')\rangle$. 
\end{itemize}
\subsection{Large $N$ dynamics at zero temperature}
\begin{figure}
\centering
\includegraphics[width=.9\columnwidth]{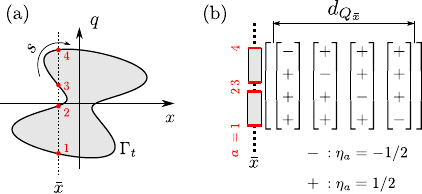}
\caption{(a)--~At time $t$ and position $\bar{x}$, the gas is in a `{split Fermi sea}' state, with Fermi points $\{q_t(s_a)\}_{a=1}^{2Q_{\bar{x}}}$ ($Q_{\bar{x}}=2$ in the figure) where $s_a, a=1,\dots,2Q_{\bar{x}}$ are the solutions of $x_t(s)=\bar{x}$. (b)--~The set ${\cal I}_{Q_{\bar{x}}}$ (here with $d_{Q_{\bar{x}}}= |{\cal I}_{Q_{\bar{x}}}| = 4$) of sequences $\eta = \{ \eta_a \}_{a=1}^{2 Q_{\bar{x}}}$ such that $\eta_a=\pm 1/2$ and $\sum_{a=1}^{2Q_{\bar{x}}}\eta_a=1$. Each sequence $\eta$ is shown as a column of the array.
}\label{fig:illustration}
\end{figure}

For the associated fermionic model, the Wigner function is
\begin{equation}
	\label{eq:wigner_function}
	W(x,q) \, = \, \frac{1}{2\pi \hbar}  \int dy \, e^{\frac{i q y}{\hbar}} \langle \hat{\Psi}^\dagger_{\rm F}(x+ \frac{y}{2}) \hat{\Psi}_{\rm F}(x-\frac{y}{2}) \rangle,
\end{equation}
and measures the phase-space fermionic occupation. In the ground state in an initial trapping potential $V(x)= V_0(x)$, it has a simple semiclassical limit reflecting the fact that all single-particle states with negative energies are occupied,
\begin{equation}
	W (x,q) \, \underset{\hbar \rightarrow 0}{=} \,  \left\{
		\begin{array}{cl}
			1/(2\pi \hbar)  &{\rm if} \quad \frac{q^2}{2} + V_0(x) <0 , \\
			0 &{\rm otherwise} .
		\end{array}
		\right.
\end{equation}
As pointed out by many authors~\cite{bettelheim2011universal,bettelheim2012quantum,kulkarni2018quantum,ruggiero2019conformal,dean2019nonequilibrium,ruggiero2020quantum}, the limit $\hbar \rightarrow 0$ is a thermodynamic limit. Indeed, the number of atoms in the cloud is $N = \int W(x,q) \, dx\, dq$, and it goes as
\begin{equation}
N  \, \sim \, 1/\hbar .
\end{equation}
Therefore, in the following, we access the large $N$ behavior of the gas by taking the limit $\hbar \rightarrow 0$. 
For simplicity, we assume that the initial potential is such that $V_0(x)<0$ in an interval $x\in [-R,R]$, so that when the gas is prepared in the ground state of $\hat{H}$, there is a single atom cloud containing $N = \frac{1}{\pi \hbar} \int_{-R}^R \sqrt{-2V_0(y)} dy$ atoms. The Wigner function evolves according to the Moyal evolution equation~\cite{moyal1949quantum,fagotti2017higher,fagotti2020locally}. Up to corrections that are subleading in $1/N \sim \hbar$, this is
\begin{equation}
	\label{eq:moyal}
	\partial_t W + q \partial_x W - (\partial_x V_1) \partial_q W  =  O(\hbar^2) .
\end{equation}
Thus, to leading order in $1/N$, the dynamics of the zero-temperature TG gas is one of an incompressible droplet in phase space that follows the classical dynamics (\ref{eq:moyal}) \cite{bettelheim2012quantum,kulkarni2018quantum,ruggiero2019conformal,dean2019nonequilibrium,ruggiero2020quantum}, see Fig.~\ref{fig:illustration}(a).

\subsection{The contour and the time-dependent WKB phase}
For our purposes, a key object is the contour of the incompressible droplet, i.e. the curve $(x,q)$ that  satisfies $\frac{q^2}{2} + V_0(x)=0$ in the initial state, and then moves along with the droplet. We parametrize the contour in the initial state as $\Gamma_0 = \{ (x_0(s) , q_0(s))  \, ; \; 0 \leq s < 2\pi  \}$. At later times, all points on the contour $\Gamma_t = \{ (x_t(s) , q_t(s))  \, ; \; 0 \leq s < 2\pi \}$ evolve like pointlike particles in the potential $V_1(x)$,
\begin{equation}\label{eq:zero-entropy-ghd}
	\frac{d}{dt} \left( \begin{array}{c} x_t(s) \\ q_t(s) \end{array} \right)  =  \left( \begin{array}{c} q_t(s) \\ - \partial_x V_1 (x_t(s))  \end{array} \right) .
\end{equation}
\indent
To the contour $\Gamma_t$, we associate a time-dependent WKB (Wentzel-Kramers-Brillouin) phase $\Phi$ along the contour, defined by the differential 
\begin{equation}
    \label{eq:differential}
	d \Phi =  \frac{1}{\hbar}  [  q \, d x  -  \varepsilon \, dt ] .
\end{equation} 
Here we are locally parametrizing the contour $\Gamma_t$ as $(x, q(x,t))$, and $\varepsilon(x,t) = q(x,t)^2/2 + V_1(x)$ is the energy of a pointlike particle at position $(x,q(x,t))$ in phase space. Notice that the cross-derivatives in Eq.~(\ref{eq:differential}) are equal thanks to the evolution equation (\ref{eq:zero-entropy-ghd})~\cite{ruggiero2019conformal}. The WKB phase $\Phi$ is only defined modulo $2\pi$ and up to an additive constant, reflecting the global $U(1)$ invariance of the model. Notice that integrating Eq.~(\ref{eq:differential}) for any fixed time gives a constant `winding number', $\int_{0}^{2\pi} d\Phi_t(s) = 2\pi N$. In the rest of the paper, we express our results using the following gauge choice for the WKB phase. At time $t=0$, we take
\begin{eqnarray}\label{eq:phi-0}
\label{eq:phi0}
	\Phi_0(s)& = &   \text{sign}(q_0(s))\int_{-R}^{x_0(s)} \sqrt{-2V_0(y)} \, dy,
\end{eqnarray}
which has a $2\pi N$-jump at the rightmost point of the cloud, $s = s^\star_0$, where $s^\star_0$ is such that $x_0(s^\star_0)= R$. Then at time $t$ we define
\begin{eqnarray}\label{eq:phi-t}
\nonumber	\Phi_t(s)& = &     \Phi_0(s) + 
 \frac{1}{\hbar} \int_0^t  \left( \frac{ ( q_\tau (s))^2 }{2}  - V_1(x_\tau(s))  \right)   d\tau  \\
&& \qquad + \;  2\pi N  \, \times \, \mathbf{1}_{[s_t^\star,s^\star_0]}(s) ,
\end{eqnarray}
where $s^\star_t$ is such that $x_t(s^\star_t)=\max_s(x_t(s))$ and $\mathbf{1}_{[s_t^\star,s^\star_0]}(s)= 1$ if $s\in[s^\star_t,s^\star_0]$ and $0$ otherwise. Our convention ensures that, at any time $t$, $\Phi_t(s)$ is a continuous function of $s$ everywhere but at $s^\star_t$, corresponding to the rightmost point of the cloud where the atom density vanishes. There, it has a $2\pi N$-jump.\\[.1cm]
\indent
Let us briefly elaborate on the parametrization of the contour $\Gamma_0$. We are free to chose the coordinate $s$ in any way we like, but we find that the most convenient choice is such that
\begin{equation}
    \label{eq:choice_s}
	\frac{d x_0(s) }{ds}  =  q_0(s)  \, \mathcal{N} , 
\end{equation}
where the constant $\mathcal{N}  = \frac{1}{\pi} \int_{-R}^R  dx/ \sqrt{-2 V(x)}$ is fixed so that $2 \int_{-R}^R  \frac{d s}{ d x_0 } d x_0  = \int_0^{2\pi}  ds  =2 \pi$.  That coordinate $s$ is interpreted as the (rescaled) time needed by an excitation originating from the left boundary of the cloud to travel to point $x_0$.
\\[.1cm]
\indent
Finally, notice that at any given time $t$ and position $x$, the contour $\Gamma_t$ intersects the vertical axis at $x$ some even number of times $2 Q_x$ (Fig.~\ref{fig:illustration}). Let $s_1, \dots, s_{2 Q_x}$ be such that $x_t(s_1) = x_t(s_2) = \dots = x_t(s_{2Q_x})\equiv x$ and $q_t(s_1) < q_t(s_2) < \dots < q_t(s_{2Q_x})$. Locally, the gas is in a state known as a `{split Fermi state}'~\cite{eliens2016general,eliens2017quantum,doyon2017large,ruggiero2020quantum} defined by the Fermi points $\{q_t(s_a)\}_{a=1}^{2Q_x}$, see Fig.~\ref{fig:illustration}(b). Such states are true local out-of-equilibrium states that clearly differ from the ground state of the gas. 

\section{One-particle density matrix}\label{sec:ana-res}
Our main result is an asymptotically exact formula for the bosonic 1PDM at time $t$, which is most conveniently expressed as a vector-matrix-vector product,
\begin{eqnarray}
\label{eq:mainresult}
 && \langle \hat{\Psi}^\dagger(x) \hat{\Psi}(x') \rangle \; \underset{\hbar \rightarrow 0 }{=} \;  \mathcal{C}^\dagger(x) \cdot  \mathcal{F}( x , x' )  \cdot \mathcal{C}(x')  \\
\nonumber && \qquad =   \sum_{ \eta \in \mathcal{I}_{Q_x} } \sum_{\eta' \in \mathcal{I}_{Q_{x'}} }   [\mathcal{C}(x)]^*_{\eta}   [ \mathcal{F}( x , x' )  ]_{\eta,\eta'}  [ \mathcal{C}(x')]_{\eta'}  .
\end{eqnarray}
Here the entries of the vectors and of the matrix are labeled by sequences  $\eta = \{\eta_a\}_{a=1}^{2Q}$ with $\eta_a = \pm 1/2$ and $\sum_{a=1}^{2Q} \eta_a = 1$, see Fig.~\ref{fig:illustration}(b). We call $\mathcal{I}_Q$ the set of such sequences, with cardinality $d_Q =|\mathcal{I}_Q| =( 2 Q)! /[ (Q-1)! (Q+1)!]$. The entries of the $d_{Q_x} \times d_{Q_{x'}}$ matrix are
\begin{eqnarray}\label{eq:F} 
[ \mathcal{F}( x , x' )  ]_{\eta,\eta'}=\frac{  \prod\limits_{a < b}^{2Q}  \left|  2 \sin \frac{s_a - s_b}{2}  \right|^{\eta_{a} \eta_{b}}   \prod\limits_{c < d}^{2Q'}  \left|  2 \sin \frac{s'_c - s'_d}{2}  \right|^{\eta'_{c} \eta'_{d}}   }{  \prod\limits_{i =1 }^{2Q}  \prod\limits_{j =1 }^{2Q'}  \left|  2 \sin \frac{s_i - s'_j}{2}  \right|^{\eta_{i} \eta'_{j}}  } \; \;
\end{eqnarray}
and the ones of the $d_{Q_x}$-dimensional vector $\mathcal{C}(x)$ are
\begin{equation}
	[ \mathcal{C}( x  )  ]_{\eta} =  \frac{ ( \frac{ G^{2} (3/2) }{\sqrt{\pi}}   )^{Q_x} }{\sqrt{2}}   \prod_{j = 1}^{2 Q_x} \left| \frac{ds_j}{dx} \right|^{\frac{1}{8}}  e^{-i\eta_j\Phi_j}\prod_{a<b}^{2Q_x}  \left| q_a - q_b  \right|^{\eta_a \eta_b},
	\label{CC}
\end{equation}
where $q_a\equiv q_t(s_a)$, $\Phi_a\equiv \Phi_t(s_a)$ and $G(\cdot)$ denotes the Barnes G-function.
\begin{figure*}[t!]
\centering
\includegraphics[width=\textwidth]{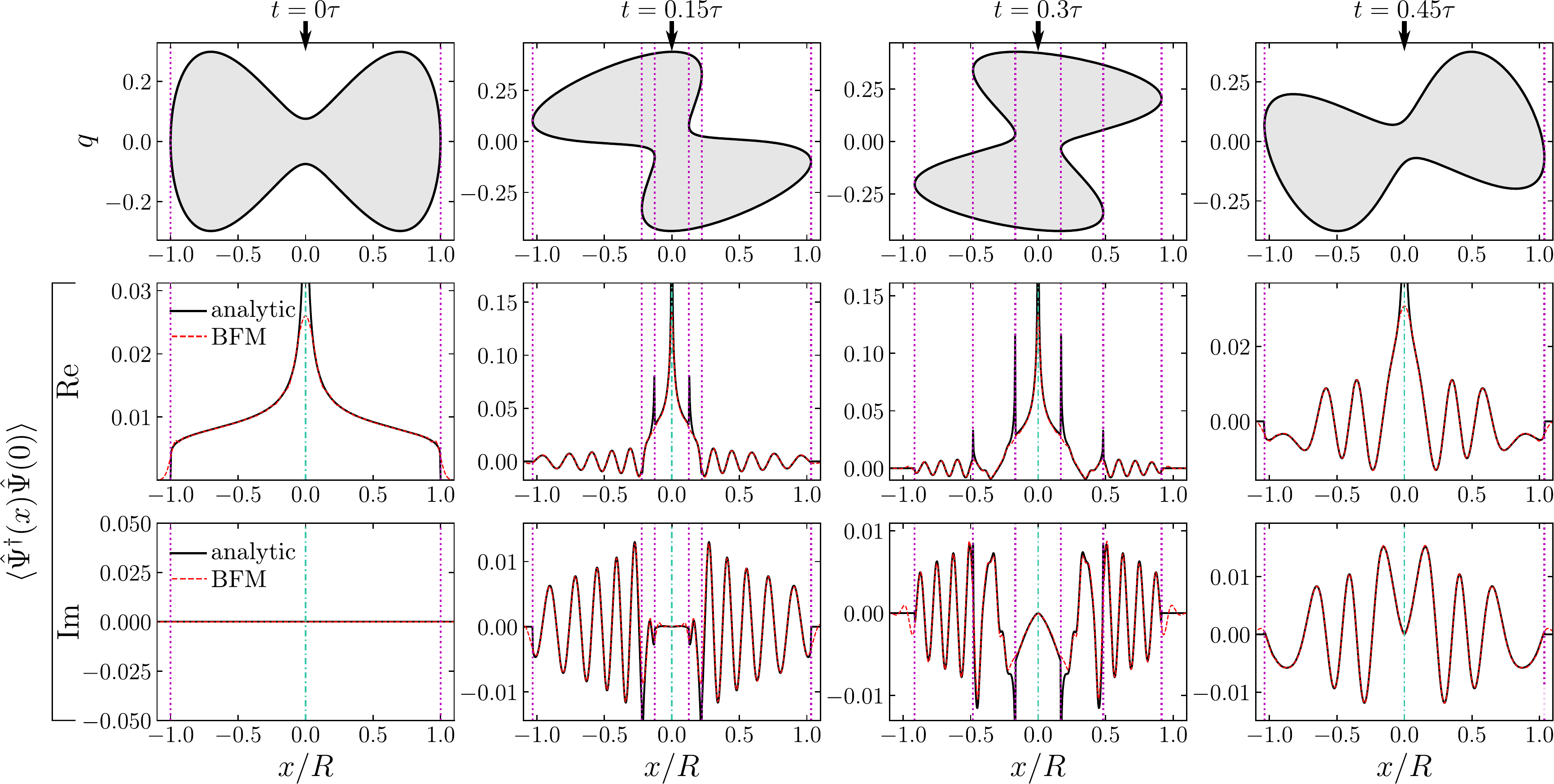}
\caption{{\it Top row}~--~Evolution of the Fermi contour $\Gamma_t$ after a quartic to quadratic trap trap quench, obtained from Eq.~\eqref{eq:zero-entropy-ghd}. {\it Bottom rows}~--~ Evolution of the corresponding bosonic 1PDM for $x'=0$. Full lines: analytical result of Eq.~\eqref{eq:mainresult}; dashed lines: BFM numerics, obtained as in Refs.~\cite{pezer2007momentum,atas2017exact}. We set $V_0=6(x/L)^4-(x/L)^2-\mu$ with $\mu=0.0028$ and $V_1=\frac{1}{2}\omega^2 x^2-\mu$ with $\omega=L^{-1}$, $L=600$ is the system's size. With this choice of parameters, the system contains $N=33$ particles. Time is expressed in units of $\tau=\frac{2\pi}{\omega}$.
}\label{fig:1pdm-split}
\vspace{.8cm}
\centering
\includegraphics[width=\textwidth]{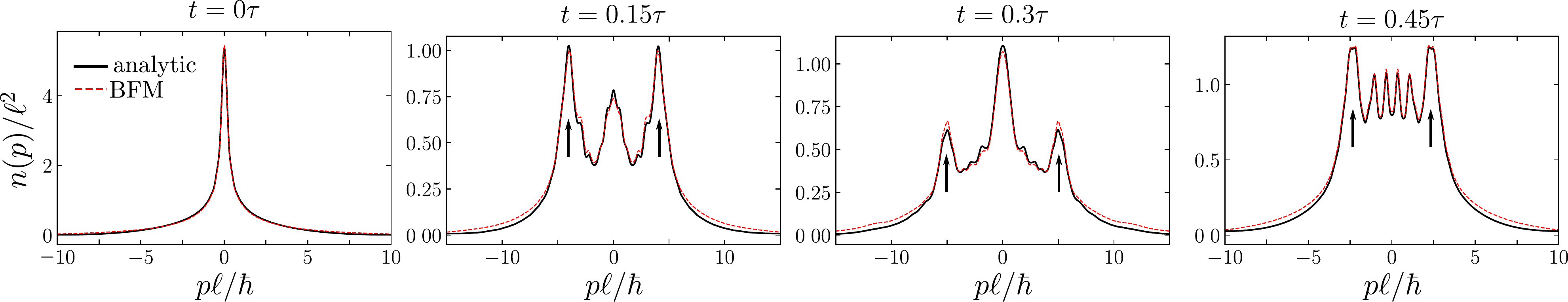}
\caption{Evolution of the MD for the quartic to quadratic trap quench of Fig.~\ref{fig:1pdm-split}. Full lines: prediction obtained from Eq.~\eqref{eq:mainresult}; dashed lines: BFM numerics. We rescaled the MD and the momenta in terms of $\ell=\sqrt{\hbar/\omega}$, $\omega$ is the frequency of $V_1$. Arrows point at the large symmetric peaks at $p\neq0$.
}\label{fig:MD}
\end{figure*}
Our result~\eqref{eq:mainresult} is valid as long as $|x-x'| \gg {\rm max} (\rho(x)^{-1}, \rho(x')^{-1})$ where $\rho(x) = \langle \hat{\Psi}^\dagger (x) \hat{\Psi}(x) \rangle$ is the local atom density. It becomes exact in the limit $N \sim 1/\hbar \rightarrow \infty$ (with positions $x$ and $x'$ fixed independently of $N$). At equilibrium ($V_1 = V_0$), it coincides with the known exact results of Refs.~\cite{forrester2003finite,brun2017one}, and with those of Ref.~\cite{ruggiero2019conformal} in the special case of a quench from harmonic to harmonic potential ---the latter case does not display split Fermi seas and is solvable by other methods~\cite{minguzzi2005exact,gangardt2004universal,brun2017one}---see  Appendix~\ref{sec:further-results}. Equation~\eqref{eq:mainresult} provides a long sought-after, and highly non-trivial, generalization of these exact results to a general out-of-equilibrium situation generated by a quench with arbitrary potentials $V_0$ and $V_1$. 
\subsection{Brief sketch of derivation of formula~\eqref{eq:mainresult}}
We have derived formula (\ref{eq:mainresult}) by applying the ideas of `quantum generalized hydrodynamics'~ \cite{ruggiero2020quantum,scopa2021exact,ruggiero2021quantum,scopa2022exact}, a recent theoretical framework that aims at describing quantum fluctuations and correlations of 1D fluids with nearly integrable dynamics (for introductions to generalized hydrodynamics see, e.g., Refs.~\cite{castro2016emergent,bertini2016transport,doyon2020lecture,alba2021generalized,bouchoule2022generalized}). The complete derivation of formula (\ref{eq:mainresult}) is technical and is deferred to Appendices \ref{sec:lowE-exp} and \ref{sec:coef}; here we sketch the main ingredients. The idea is that long wavelength quantum fluctuations in the fluid are encoded as small deformations $q_t(s)\to q_t(s) +\delta q_t(s)$ along the contour $\Gamma_t$, and promoted to quantum operators $\{\delta\hat{q}_a\}_{a=1}^{2Q}$ measuring the excess density of particles around the position $x\equiv x_t(s_a)$ due to the formation of a particle-hole pair~\cite{ruggiero2020quantum,moller2022bridging}. The effective field theory that captures the long-distance correlations of the operators $\{\delta\hat{q}_a\}_{a=1}^{2Q}$ is a Gaussian bosonic theory, similar to a Luttinger liquid theory~\cite{giamarchi2003quantum,tsvelik2007quantum}. The atom annihilation operator $\hat\Psi(x)$ in the microscopic model~\eqref{eq:ham} is then formally expanded in a basis of operators $\hat{\cal O}(x)$ in the effective field theory,
\be\label{eq:psi-lowE}
\hat\Psi(x)\approx {\cal C}(x) \cdot \hat{\cal O}(x)=\sum_{\eta\in{\cal I}_{Q}} [{\cal C}(x)]_\eta [\hat{\cal O}(x)]_\eta\ ,
\ee
where ${\cal C}(x)$ is an array of non-universal numerical coefficients \eqref{CC}, whose calculation is detailed in Appendix~\ref{sec:coef}. The connection between $\{\delta\hat{q}_a\}$ and $\hat{\cal O}(x)$ is established via bosonization arguments~\cite{vlijm2016correlations,eliens2016general,eliens2017quantum,ruggiero2020quantum}, according to which the excess density of quasi-particles near the $a^{\rm th}$ Fermi point is related to the derivative of a chiral boson operator $\hat{\varphi}(s)$,  
\be
\delta\hat{q}_t(s_a) =\hbar \partial \hat\varphi(s_a).
\ee
 Then the normal-ordered exponentials of the boson field, $
[\hat{\cal O}(x)]_\eta=\prod_{a=1}^{2Q_x}:e^{-i \eta_a \hat\varphi(s_a)}:$,
correspond to all the possible deformations $\eta\in{\cal I}_Q$ of the split Fermi sea with the lowest possible scaling dimension of $\hat\Psi$, see Appendix~\ref{sec:lowE-exp} for more details. \\

As pointed out in Refs.~\cite{ruggiero2020quantum,ruggiero2021quantum,scopa2021exact,scopa2022exact}, the Hamiltonian governing the dynamics of these quantum fluctuations has the quadratic form $\hat{H}[\Gamma_t]=(\pi\hbar/{\cal N})\int ds \ (\partial_s\hat\varphi)^2$ and it is sensitive only to the comoving coordinate $s$ along the contour of the phase-space droplet $W(x,q)$. This, together with the convenient choice of parametrization (\ref{eq:choice_s}) of the contour in the initial state, leads to the following simple form for the equal-time boson-boson function~\cite{ruggiero2020quantum,ruggiero2021quantum}:
\be\label{eq:propagator} 
\langle \hat\varphi(s_a)\hat\varphi(s_b)\rangle=-\log\left( 2i\sin\frac{s_a-s_b}{2}\right).
\ee
Our formula~\eqref{eq:mainresult} is then obtained by applying Wick's theorem for the field $\hat{\varphi}(s)$.
\subsection{Numerical check of formula (\ref{eq:mainresult})}
In Fig.~\ref{fig:1pdm-split} we compare the analytical result (\ref{eq:mainresult}) to a numerical calculation of the 1PDM for $N=33$, performed using time-dependent BFM, see e.g.  Refs.~\cite{pezer2007momentum,atas2017exact}. We study a quench from a double-well (quartic) potential $V_0(x)$ to a simple-well (quadratic) potential $V_1(x)$, see the caption of Fig.~\ref{fig:1pdm-split} for specific parameters. We find that the agreement is excellent, with most of the asymptotic features of our analytical formula present already for $N=33$. Our formula for the 1PDM has a UV divergence at $x=x'$ (dash-dotted lines) reminiscent of the standard Luttinger liquid result $\langle \hat{\Psi}^\dagger(x) \Psi(x') \rangle \propto|x-x'|^{-1/2}$, present also at equilibrium~\cite{forrester2003finite,papenbrock2003ground,gangardt2004universal}. In the microscopic description of the TG gas there is no divergence, since $\langle\hat\Psi^\dagger(x)\hat\Psi(x')\rangle\to\rho(x)$  when $x' \to x$. There is no contradiction since our asymptotic formula is obtained from a large-scale quantum hydrodynamic approach, so it does not apply at distances $|x-x'|$ smaller than the interparticle distance $\sim \rho(x)^{-1}$.  Additional `spikes' emerge during the time evolution, at the positions of the `turning points' of the contour $\Gamma_t$, i.e. where the number of local Fermi seas changes from one to two (dotted lines). The origin of these short-distance spikes is similar to the divergence at $x=x'$: they are inherent to the large-$N$ field-theoretic approach we are following, although they are absent from the microscopic system. Again, this reflects the fact that our asymptotic formula does not apply on distances smaller than $\sim \rho(x)^{-1}$ near the positions of the turning point. In practice, these spikes can simply be removed via local linear interpolation (as discussed in Appendix~\ref{sec:regularization}).
\subsection{Application to the calculation of the MD}
Finally, we compute the out-of-equilibrium MD of the 1D TG gas, by taking the double Fourier transform \eqref{MD1} of our formula (\ref{eq:mainresult}). In Fig.~\ref{fig:MD}, we report our result for the MD corresponding to the 1PDM in Fig.~\ref{fig:1pdm-split} (with  `spikes' removed by local linear interpolation), compared with BFM numerics~\cite{pezer2007momentum,atas2017exact}. The agreement is excellent on a wide range of momenta. Small deviations are observed on the large momentum tails of the MD since our formula does not capture the short-distance behavior of the 1PDM. This inaccuracy can be reduced  improving the UV regularization, or by combining our approach with Tan's contact physics \cite{minguzzi2002high,olshanii2003short,rigol2004universal,vignolo2013universal,decamp2016high,yao2018tan,bouchoule2021breakdown} and local density approximation~(as done for instance in Ref.~\cite{caux2019hydrodynamics}).\\
\indent

Physically, we observe the dynamical appearance of two large symmetric peaks at non-zero momenta in the MD (arrows in Fig.~\ref{fig:MD}), which are a consequence of the oscillating tails of the 1PDM (Fig.~\ref{fig:1pdm-split}). Interestingly, we note that the experimentally measured MD in the original QNC experiment~\cite{kinoshita2006quantum} also displayed such peaks~(although a direct comparison with the data of Ref.~\cite{kinoshita2006quantum} is not possible, as the quenching protocol is different: dynamics is imparted by a Bragg pulse as opposed to a quench of the trapping potential). These peaks are a fundamental qualitative non-equilibrium feature of the gas, which essentially reflect the fact that the cloud is made of a fraction of atoms going to the left, and the same fraction of atoms going to the right. In addition to these large peaks at non-zero momenta, we observe the formation of intriguing smaller structures in the MD, which evolve into smaller peaks or oscillations, e.g. at $t=0.45\tau$; so far we have not found a simple explanation for these smaller oscillations.

\section{Conclusion}
Motivated by the long-standing problem of the computation of the MD in strongly correlated ultracold gases, especially in 1D Bose gases, we derived an analytical formula ---Eq.~(\ref{eq:mainresult})--- for the 1PDM of the out-of-equilibrium TG  gas at large $N$, applicable for a gas initially prepared in its ground state in a trapping potential $V_0(x)$, with dynamics imparted by a quench $V_0(x) \rightarrow V_1(x)$. This result extends, in a very non-trivial way, some milestone results  {about the 1PDM of the TG gas} that were obtained only at equilibrium~\cite{lenard1964momentum,vaidya1979one,forrester2003finite,gangardt2004universal,brun2017one} or in the very special case of a frequency quench in a harmonic potential~\cite{minguzzi2005exact}. By comparing with BFM numerics, we have established that our formula provides a quantitatively accurate and reliable method to compute the MD in a wide range of momenta. It captures dynamical features of the MD observed in experiments that so far remained unexplained.

\begin{acknowledgements}
PC and SS acknowledge support from ERC under Consolidator grant~No.~771536 (NEMO). JD acknowledges support from CNRS International Emerging Actions
under the QuDOD grant, and from the Agence National de la Recherche through Grants No.~ANR-20-CE30-0017-01  (QUADY) and No.~ANR-18-CE40-0033  (DIMERS). We are grateful to Alvise Bastianello, Isabelle Bouchoule, Benjamin Doyon, Jacopo de~Nardis, Maurizio Fagotti, and Jean-Marie St\'ephan for useful discussions.
\end{acknowledgements}

\appendix
\input{Appendix.tex}

\bibliography{momentum_dist}
\end{document}

%% file: Appendix.tex
\section{Low energy expansion of the bosonic field}\label{sec:lowE-exp}
In this Appendix, we discuss the derivation of the low-energy expansion of the bosonic field in Eq.~\eqref{eq:psi-lowE}. For a better exposition, we briefly recall the strategy for an equilibrium configuration before considering the generic case out-of-equilibrium. We refer to, e.g., Refs.~\cite{giamarchi2003quantum,cazalilla2004bosonizing,brun2018inhomogeneous,scopa2020one} for a detailed derivation of the equilibrium results which follow. \\

At equilibrium, it is well known that the bosonic field allows for a low-energy expansion in terms of operators of an asymptotic field theory, namely
\be\label{bosonization}
\hat\Psi(x)=B(x) \ \hat{O}(x) +\text{less relevant operators}
\ee
where $B(x)$ is a dimensionful non-universal coefficient and we defined the vertex operator as
\be\label{eq:vertex}
\hat{O}(x)=\prod_{a=1,2} \left|\frac{d x_t(s_a)}{ds}\right|^{-\Delta/2} \ :e^{-\frac{i}{2}\hat\varphi(s_1)}: :e^{-\frac{i}{2}\hat\varphi(s_2)}: \ .
\ee
Here, $\hat\varphi(s)\in\mathbb{R}/(2\pi\mathbb{Z})$ is a compact chiral bosonic field living along the contour and parametrizing the chiral density fluctuations around the Fermi points as $\delta\hat{q}(s)=\hbar \partial_s\hat\varphi(s)$, see e.g. Refs.~\cite{ruggiero2021quantum,cazalilla2004bosonizing} for further details. The coordinates $s_1,s_2$ denote the positions along the contour of the Fermi points $q(s_1)=-q(s_2)$ satisfying $x\equiv x(s_1)=x(s_2)$ and $\Delta=1/4$ is the scaling dimension of the vertex operator. Notice that, in writing Eq.~\eqref{bosonization}, we considered only the low-energy excitations corresponding to a change in the particle number $N\to N-1$ and we neglected Umklapp processes which would contribute to the low-energy expansion \eqref{bosonization} with vertex operators of higher scaling dimensions.\\

\begin{figure}[t!]
\centering
\includegraphics[width=.9\columnwidth]{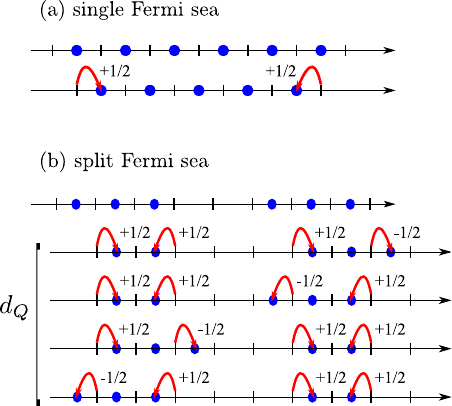}
\caption{Microscopic configuration of momenta for the Tonks-Girardeau gas before and after the removal of a particle: (a) {\it single Fermi sea} ($Q=1$)~--~The particle loss leads to the change of parity sector of the quantized momenta. Each Fermi point is ``pushed inwards" of the amount $+1/2$ as encoded by the action of the Luttinger fields $\hat\varphi(s)$. (b) {\it Split Fermi sea}~--~The single particle loss can be realized in $d_Q$ different configurations, each corresponding to a set of  values $\eta=\{\eta_a=\pm1/2\}$ to assign to each Fermi point $q_a$ depending on wheter $q_a$ is moved inwards ($\eta_a=+1/2$) or outwards ($\eta_a=-1/2$) with respect to the initial configuration. }\label{fig:eta}
\end{figure}

At equilibrium, one finds the symmetry of Fermi points $s_1=2\pi - s_2$, thanks to which the total WKB phase $\Phi_0(s_1)+\Phi_0(s_2)$ simply vanishes (cf.~Eq.~\eqref{eq:phi-t}). In the case out-of-equilibrium with a single Fermi sea ($Q=1$), we find a similar expression but the condition on the WKB phases is no longer valid. Therefore, Eq.~\eqref{bosonization} modifies as
\be\label{bosonization-1FS}
\hat\Psi(x)\approx {\cal C}(x) \ \hat{\cal O}(x)
\ee
where we defined
\be
{\cal C}(x)=B(x) \prod_{a=1,2} \left|\frac{d x_t(s_a)}{ds}\right|^{-1/8}  \exp\left(-\frac{i}{2}\Phi_t(s_a)\right)
\ee
and 
\be
\hat{\cal O}(x)=\ : \ e^{-\frac{i}{2}\hat\varphi(s_1)} \ :: \ e^{-\frac{i}{2}\hat\varphi(s_2)} \ : \ .
\ee
We observe that, in our convention, the action of the fields $\hat\varphi(s_{a=1,2})$ is to ``push inwards" the Fermi contour of an amount $+1/2$ such that the combined action of the two fields describes the loss of one atom operated by $\hat\Psi(x)$ and the consequent change of parity in the quantization of the modes, see Fig.~\ref{fig:eta}(a).\\

At this point, in generalizing the expression \eqref{bosonization-1FS} to an arbitrary number $Q$ of Fermi seas, there are $d_Q$ possible configurations of the split Fermi sea in which a particle can be removed, see Fig.~\ref{fig:eta}(b). We denote each of these configurations with a $2Q$-dimensional vector $\eta$ satisfying
\be
\eta_a=\pm 1/2, \quad \sum_{a=1}^{2Q} \eta_a=1
\ee
such that the total action of the fields $\hat\varphi(s_a)$ correctly reproduce the action of the operator $\hat\Psi(x)$. Since each configuration $\eta$ contributes to the low-energy expansion of $\hat\Psi$ with equal scaling dimension $\Delta=1/4$, a sum over configuration is required and Eq.~\eqref{bosonization-1FS} becomes
\be\label{bosonization-QFS}
\hat\Psi(x)\approx {\cal C}(x)\cdot \hat{\cal O}(x)= \sum_{\eta\in {\cal I}_Q} [{\cal C}(x)]_\eta [\hat{\cal O}(x)]_\eta
\ee
where
\be\label{coef-QFS}
[{\cal C}(x)]_\eta=B_\eta(x) \prod_{a=1}^{2Q} \left|\frac{d x_t(s_a)}{ds}\right|^{-1/8} \ e^{-i\eta_a\Phi_t(s_a)}  
\ee
and
\be
[\hat{\cal O}(x)]_\eta=\prod_{a=1}^{2Q} :e^{-i\eta_a\hat\varphi(s_a)}: \ .
\ee
Notice that in the case $Q=1$, we obtain a single configuration $\eta=\{+1/2,+1/2\}$ and Eq.~\eqref{bosonization-QFS} reduces to \eqref{bosonization-1FS}. The calculation of the (dimensionful) non-universal coefficient $B_\eta(x)$ appearing in Eq.~\eqref{coef-QFS} is discussed below.

\section{Calculation of the non-universal amplitudes}\label{sec:coef}
As previously discussed in Refs.~\cite{scopa2020one,brun2018inhomogeneous,shashi2012exact,shashi2011nonuniversal}, the non-universal coefficient $B_\eta$ can extracted from the field form factor of the microscopic model at finite $N,L$ as
\be\label{coef}
\begin{split}
&B_\eta(x)=\lim_{N,L\to \infty} \left(\frac{L}{2\pi}\right)^{Q/4} \times \\
& \frac{\left|\langle\{q^{(\eta)}_i\}_{i=1}^{N-1}|\hat\Psi(0)|\{k_j\}_{j=1}^N\rangle\right|}{\sqrt{\langle\{q^{(\eta)}_i\}_{i=1}^{N-1}|\{q^{(\eta)}_i\}_{i=1}^{N-1}\rangle}\sqrt{\langle\{k_j\}_{j=1}^{N}|\{k_j\}_{j=1}^{N}\rangle}}
\end{split}
\ee
where the limit $N,L\to\infty$ is taken with fixed ratio $N/L=\rho(x)$, with $\rho(x)$ being the particle density at position $x$.  The state $|\{k_i\}\rangle$ is a reference state for the microscopic model while $|\{q^{(\eta)}_i\}\rangle$ is an excited state depending on the particular configuration $\eta$ which is considered. For arbitrary values of momenta of the in- ($I=\{k_i\}_{i=1}^N$) and out- ($J_\eta=\{q^{(\eta)}_i\}_{i=1}^{N-1}$) states, the field form factor in Eq.~\eqref{coef} for the Tonks-Girardeau gas is \cite{slavnov1989calculation}
\be\label{FF}
\begin{split}
 {\cal G}(I|J_\eta)&\equiv\frac{|\langle J_\eta|\hat\Psi(0)|I\rangle|}{\sqrt{\langle J_\eta|J_\eta\rangle}\sqrt{\langle I|I\rangle}}\\
&=\frac{2^{N-1}}{L^{N-\frac{1}{2}}}\frac{\prod\limits_{1\leq a<b\leq N}|k_a-k_b| \prod\limits_{1\leq c<d\leq N-1} |q^{(\eta)}_c-q^{(\eta)}_d|}{\prod\limits_{i=1}^N \prod\limits_{j=1}^{N-1}|k_i-q^{(\eta)}_j|}
\end{split}\ee 
with $I\subset \frac{2\pi}{L}(\mathbb{Z}+1/2)$ and $J_\eta\subset \frac{2\pi}{L}\mathbb{Z}$, assuming even $N$. In detail:\\

\subsubsection{ Single Fermi sea ($Q=1$)}
Let $|\{k_j\}\rangle$ be the ground state of the microscopic model having $N$ particles, specified by the set of momenta
\be\label{set1-eq}
k_j=\frac{2\pi}{L}\left(-\frac{N+1}{2}+j\right), \quad j=1,\dots,N
\ee
and $|\{q_i\}\rangle$ the excited state obtained by removing a particle from the ground state and specified by the momenta
\be\label{set2-eq}
q_i =\frac{2\pi}{L}\left(-\frac{N}{2}+i\right), \quad i=1,\dots,N-1.
\ee
For out-of-equilibrium configurations, we notice that a uniform boost $\Lambda$ of the momenta in \eqref{set1-eq} and \eqref{set2-eq} does not modify the value of $B$ (cf.~Eqs.~\eqref{coef} and \eqref{FF}). By evaluating Eq.~\eqref{FF} with the sets of momenta in \eqref{set1-eq} and \eqref{set2-eq}, we obtain
\be
 {\cal G}(\{q_i\}_{i=1}^{N-1}|\{k_j\}_{j=1}^N)=L^{-1/2} \frac{G^2(3/2)G(N)G(N+1)}{G^2(N+1/2)},
\ee
and by expanding the Barnes G-function for large $N$ as $\frac{G(N)G(N+1)}{G^2(N+1/2)}\sim N^{1/4}$, we recover the known result (see e.g. Refs.~\cite{lenard1964momentum,widom1973toeplitz})
\be\label{coef-equilibrium}
B(x)= \frac{G^2(3/2)}{(2\pi)^{1/4}}\ (\rho(x))^{1/4}.
\ee
By combining the scaling dimensions of the non-universal amplitude $B\propto \rho^{1/4}$ with $\Delta=1/4$ of the vertex operator in \eqref{bosonization-1FS}, we recover the correct scaling dimension $\Delta_\Psi=1/2$ of the bosonic field.\\

\subsubsection{Split Fermi sea}
We now turn to the generic out-of-equilibrium situation. In this case, typical states have the form of a split Fermi sea with boundaries $\{k_a\}_{a=1}^{2Q}$ such that 
\be
\sum_{a=0}^{Q-1} \frac{k_{2a+2}-k_{2a+1}}{2\pi}=N/L=\rho(x).
\ee
For even $N$, the quantized momenta populating the split Fermi sea are obtained by the set
\be
I=\frac{2\pi}{L}(\mathbb{Z}+\frac{1}{2})\cap\left([k_1,k_2]\cup\dots\cup[k_{2Q-1},k_{2Q}]\right)
\ee
while, after removing a particle, we have the configuration
\begin{widetext}
\be
J_\eta=\frac{2\pi}{L}\mathbb{Z}\cap\left([k_1+\eta_1,k_2-\eta_2]\cup \dots\cup[k_{2Q-1}+\eta_{2Q-1},k_{2Q}-\eta_{2Q}]\right).
\ee
\end{widetext}
For these sets, the form factor \eqref{FF} at large $N$ and for $k_a\gg 1$ is
\be
{\cal G}(I|J_\eta)\simeq \sqrt{\frac{\pi}{L}}\left(\frac{L}{2\pi}\right)^{\frac{2-Q}{4}}\left(\frac{G^2(3/2)}{\sqrt{\pi}}\right)^Q \ \prod_{a<b}^{2Q}|k_a-k_b|^{\eta_a\eta_b}
\ee
leading to the non-universal coefficient
\be\label{coef-FS}
B_\eta(x)= \frac{ ( \frac{ G^{2} (3/2) }{\sqrt{\pi}}   )^{Q} }{\sqrt{2}}   \prod_{a<b}^{2Q}  \left| k_a - k_b  \right|^{\eta_a \eta_b}.
\ee
One can easily check that for $Q=1$ this expression reduces to Eq.~\eqref{coef-equilibrium}. Plugging Eq.~\eqref{coef-FS} into Eq.~\eqref{coef-QFS}, we recover the expression for the coefficient $[{\cal C}(x)]_\eta$ appearing in Eq.~\eqref{eq:psi-lowE}.

\section{Further results for the 1PDM}\label{sec:further-results}
In this Appendix, we provide further results and analytical checks of our asymptotic formula in Eq.~\eqref{eq:mainresult} of the main text.
\subsection{Equilibrium limit}
We first show how Eq.~\eqref{eq:mainresult} reduces to the known asymptotic result for the 1PDM at equilibrium, previously derived in Ref.~\cite{brun2017one}. Using the results of Appendix~\ref{sec:lowE-exp} and Appendix~\ref{sec:coef}, we can write the 1PDM as
\be\begin{split}
&\langle \hat\Psi^\dagger(x)\hat\Psi(x')\rangle=B(x)B(x') \ \langle\hat{O}(x)\hat{O}(x')\rangle\\[4pt]
&=\quad \frac{G^4(3/2)}{\sqrt{2\pi}} \rho(x)^{1/4}\ \rho(x')^{1/4}  \prod_{a=1}^4 \left|\frac{dx_0(s_a)}{ds}\right|^{-\frac{1}{8}} \\
&\qquad\times\langle :e^{\frac{i}{2}\hat\varphi(s_1)}::e^{\frac{i}{2}\hat\varphi(s_2)}::e^{-\frac{i}{2}\hat\varphi(s_3)}::e^{-\frac{i}{2}\hat\varphi(s_4)}:\rangle
\end{split}\ee
where $s_1$, $s_2$ denote the Fermi points $x\equiv x_0(s_1)=x_0(s_2)$ and $s_3$, $s_4$ denote those satisfying $x'\equiv x_0(s_3)=x_0(s_4)$. At $t=0$ (i.e., for an equilibrium configuration), it is easy to see that
\be
\left|\frac{dx_0(s_a)}{ds}\right|=\frac{1}{{\cal N}\rho(x_0(s_a))}.
\ee
Using Eq.~(${\color{red}16}$)  and the relations $s_1\equiv s_x=2\pi-s_2$ and $s_4\equiv s_{x'}=2\pi-s_3$, after simple algebra, one obtains 
\be
\langle \hat\Psi^\dagger(x)\hat\Psi(x')\rangle=\frac{\left(\frac{G^4(3/2)}{\sqrt{2\pi}}\right)}{\sqrt{2{\cal N}}} \frac{|\sin(s_x)|^{\frac{1}{4}}|\sin(s_{x'})|^{\frac{1}{4}}}{|\sin(\frac{s_x-s_{x'}}{2})|^{\frac{1}{2}}|\sin(\frac{s_x+s_{x'}}{2})|^{\frac{1}{2}}}
\ee
recovering the result first obtained in Ref.~\cite{brun2017one}.
\subsection{Dynamics of the 1PDM in harmonic traps}
We now discuss the case of the Tonks-Girardeau gas in a harmonic potential $V_0(x)=\frac{1}{2}\omega x^2 -\mu$ and subject to a quantum quench where the trap's frequency suddenly changes from $\omega$ to $\Omega$.  For this specific setup, analytical results for the 1PDM have been obtained in Refs.~\cite{minguzzi2005exact,ruggiero2019conformal} exploiting the known exact solution for the single-particle Schr\"odinger equation (see e.g.~Refs.~\cite{pinney1950nonlinear,lewis1969exact}). In the following, we show how our asymptotic formula~\eqref{eq:mainresult} reduces to the known result in this limiting case.\\

\begin{figure}[!t]
\centering
	\includegraphics[width=.9\columnwidth]{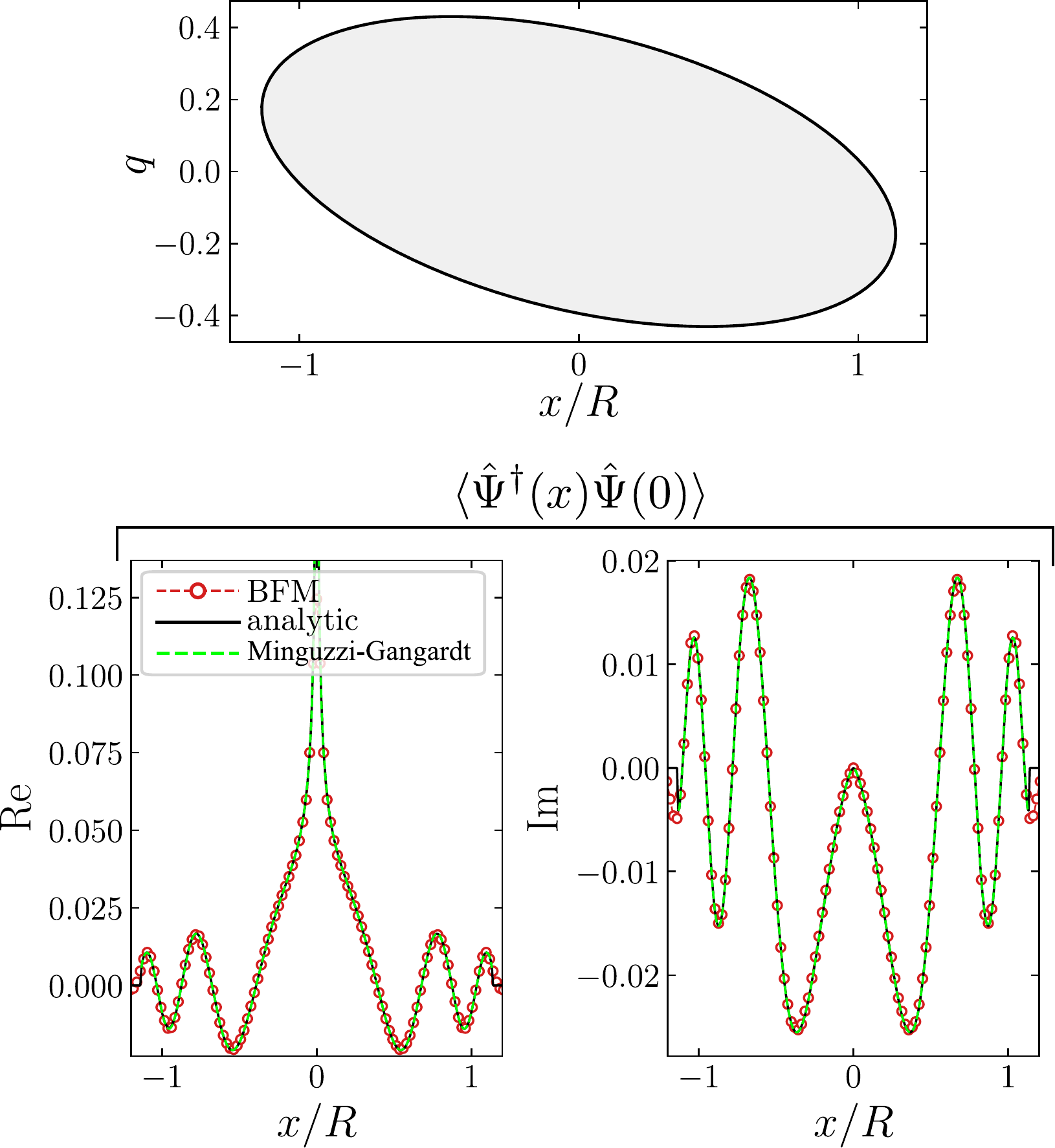}
\caption{{\it Top panel}~--~Snapshot of the Fermi contour $\Gamma_t$ during a quantum quench in the harmonic trap's frequency $\omega\to\Omega$, obtained from the solution of Eq.~(${\color{red}7}$) of the main text. {\it Bottom panels}~--~ Real and imaginary part of the time-evolved 1PDM for $x'=0$. We show the analytical results in Eq.~\eqref{Harmonic} (full line) and Eq.~\eqref{eq:Minguzzi-Gangardt} (dashed line) against BFM numerics (markers) obtained with the method discussed in Ref.~\cite{pezer2007momentum,atas2017exact}. In the figures, we set $V_0=\frac{1}{2}\omega^2x^2-\mu$ with $\omega=2/L$, $\mu=0.1$ and $V_1=\frac{1}{2}\Omega^2 x^2-\mu$ with $\Omega/\omega=2$, $L=600$ is the size of the system. With this choice of parameters, the system contains $N=30$ particles. Time is set to $t=0.225\tau$, in units of $\tau=\frac{2\pi}{\Omega}$.}\label{fig:harmonic}
\end{figure}

Applying our formalism, one can easily see that the bosonic 1PDM during a harmonic-to-harmonic trap quench has the form (hereafter $\hbar=1$)
\be\begin{split}\label{Harmonic}
&\langle\hat\Psi^\dagger(x)\hat\Psi(x')\rangle=\frac{G^4(3/2)}{\sqrt{2\pi}} \rho(x)^{1/4}\ \rho(x')^{1/4}\\
&\quad\times  \prod_{a=1}^4 \left|\frac{dx_t(s_a)}{ds}\right|^{-\frac{1}{8}} e^{\frac{i}{2}[\Phi_t(s_1)+\Phi_t(s_2)-\Phi_t(s_3)-\Phi_t(s_4)]}\\
&\quad \times \frac{|\sin(\frac{s_1-s_2}{2})|^\frac{1}{4}|\sin(\frac{s_3-s_4}{2})|^\frac{1}{4}}{|\sin(\frac{s_1-s_3}{2})|^\frac{1}{4}|\sin(\frac{s_1-s_4}{2})|^\frac{1}{4}|\sin(\frac{s_2-s_3}{2})|^\frac{1}{4}|\sin(\frac{s_2-s_4}{2})|^\frac{1}{4}}
\end{split}\ee
since the problem is characterized by a single Fermi sea for any position $x$ and time $t$. Here, $s_1$, $s_2$ denote the Fermi points $x\equiv x_0(s_1)=x_0(s_2)$ and $s_3$, $s_4$ are obtained from $x'\equiv x_0(s_3)=x_0(s_4)$.
This expression can be further simplified by employing the parametrization of the initial contour
\be
\Gamma_0:\quad (x_0(s),q_0(s))=(-R\cos(s),R\omega\sin(s))
\ee
where $R=\sqrt{2\mu}/\omega$ is the size of the cloud at $t=0$. The single-particle evolution in the harmonic trap is
\be
\left(\begin{array}{cc}
x_t(s)\\[4pt]
q_t(s)
\end{array}\right) =\left(\begin{array}{cc}
\cos(\Omega t) & \sin(\Omega t)/\Omega\\[4pt]
-\Omega\sin(\Omega t) & \cos(\Omega t) 
\end{array}\right)\left(\begin{array}{c}
x_0(s) \\[4pt]
q_0(s)
\end{array}\right)
\ee
from which one can obtain exact expressions for the jacobian $dx_t(s)/ds$ appearing in Eq.~\eqref{Harmonic}. The WKB phase is
\be\begin{split}
&\Phi_t(s)=\Phi_0(s)+(2\pi N)\mathbf{1}_{[s^\star_t,\pi]}(s)\\
&\quad-q_0(s)x_0(s)\sin^2(\Omega t)+\frac{(q_0(s)^2-x_0(s)^2)}{4\Omega}\sin(2\Omega t)
\end{split}\ee
where $s_t^\star$ satisfies $x_t(s^\star)=\max_s(x_t(s))$ and 
\be
\Phi_0(s)=\begin{cases}
f(s), \quad \text{if $s\in [0,\pi)$;}\\[4pt]
-2\pi N + f(s), \quad \text{if $s\in(\pi,2\pi]$;}
\end{cases}
\ee
with
\be
 f(s)\equiv\int_0^s ds \frac{dx_0(s)}{ds} \ q_0(s)=\frac{\omega R^2}{2}(s-\cos(s)\sin(s)).
\ee
By plugging these results in Eq.~\eqref{Harmonic}, we obtain a closed expression for the 1PDM which is showed in Fig.~\ref{fig:harmonic}. Notice that, in the quantum generalized hydrodynamics framework, correlations are expressed in terms of those in the initial state via Eq.~\eqref{eq:zero-entropy-ghd}. In Refs.~\cite{minguzzi2005exact,ruggiero2019conformal}, due to the exact solvability of the model, the isothermal coordinates at position $x$ and time $t$ can be written as a function of time
\be\label{iso-coo-harm}
s_\pm(x,t)=\pi \pm \arccos\left(\frac{x}{R b(t)}\right)
\ee
with $b(t)=\sqrt{1+(\omega^2-\Omega^2)\sin^2(\Omega t)/\Omega^2}$, resulting in the expression for the bosonic 1PDM
\be\label{eq:Minguzzi-Gangardt}
\langle\hat\Psi^\dagger(x)\hat\Psi(x')\rangle=\frac{G^4(3/2)}{\sqrt{2\pi}}  \frac{e^{-i\frac{\dot{b}(t)}{2b(t)}(x^2-{x'}^2)}}{\sqrt{b(t)}} \frac{\rho(x)^{\frac{1}{4}}\ \rho(x')^{\frac{1}{4}}}{\left|(x-x')/b(t)\right|^\frac{1}{2}},
\ee
first derived in Ref.~\cite{minguzzi2005exact} by Minguzzi and Gangardt. One can easily show that this result is obtained from Eq.~\eqref{eq:Minguzzi-Gangardt} using the coordinate \eqref{iso-coo-harm} and adding the phase $[\Phi(s_+(t))+\Phi(s_-(t))]/2$, see Ref.~\cite{ruggiero2019conformal} for the details of this calculation.\\

 In Fig.~\ref{fig:harmonic}, the analytical predictions for the 1PDM in Eq.~\eqref{Harmonic} and \eqref{eq:Minguzzi-Gangardt} are compared with time-dependent BFM numerics, showing excellent agreement.\\

\subsection{An example with split Fermi seas}
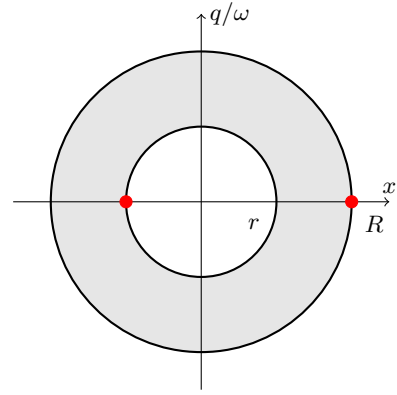
\begin{figure}[t!]
	\begin{tikzpicture}
		\filldraw[fill=ggray,draw=black,thick] (0,0) circle (2cm);
		\filldraw[fill=white,draw=black,thick] (0,0) circle (1cm);
		\draw[->] (-2.5,0) -- (2.5,0) node[above]{$x$};
		\draw[->] (0,-2.5) -- (0,2.5) node[right]{$q/\omega$};
		\draw (2.3,-0.3) node{$R$};
		\draw (0.7,-0.3) node{$r$};
		\filldraw[red] (2,0) circle (0.8mm);
		\filldraw[red] (-1,0) circle (0.8mm);
	\end{tikzpicture}	
	\caption{Illustration of the {\it ring state} in phase space, corresponding to an excited state of $N \gg 1$ hard-core particles in a harmonic potential, where the single-particle orbitals of the harmonic oscillator are filled from level $M$ to $M+N-1$. The aspect ratio is $R/r = \sqrt{1+N/M}$. The red dots indicate the point $s=\pi$ according to the parametrization below. At these points, the WKB phases undergo a discontinuity as commented in the text.}
	\label{fig:ring}
\end{figure}

In this subsection, we provide an explicit example of calculation of the 1PDM for a configuration of the Wigner function $W(x,q)$ containing split Fermi seas. Specifically, we consider the {\it ring state} depicted in Fig.~\ref{fig:ring}, obtained as excited state of a hard-core quantum gas in a harmonic trap of frequency $\omega$  with $N\gg 1$ particles filling the orbitals from $M$ to $M+N-1$.
For this state, we find a pair of Fermi contours having opposite chirality, which we denote as $\Gamma_i$ (inner) and $\Gamma_o$ (outer), respectively. A convenient parametrization for these curves is given by
\begin{eqnarray}
\Gamma_o:\quad	( x_o (s) , q_o (s) ) &=&  \left( - R \cos s , \,  \omega  R \sin s    \right) \\
\Gamma_i:\quad	( x_i (s) , q_i (s) ) &=&  \left(  r \cos s , \,  \omega  r  \sin s    \right) ,
\end{eqnarray}
where $R = \sqrt{ \frac{2 \hbar}{\omega} (N+M)} $, and $r = \sqrt{ \frac{2 \hbar}{\omega} M}$. 
For each contour, one finds the WKB phase
\begin{eqnarray}
	\Phi_{o} (s)  &=& (N+M) ( s- \frac{1}{2} \sin (2s) ), \\
	\Phi_i (s)  &=& M ( -s+ \frac{1}{2} \sin (2s) )  .
\end{eqnarray}

These phases undergo a discontinuity for $s=\pi$ (red dots in Fig.~\ref{fig:ring}), where $\Phi_o$ jumps by $2\pi (N+M)$ and  $\Phi_i$ jumps by $-2\pi M$.\\

In the region with a split Fermi sea, i.e., for $-r<x<r$, we label the four Fermi points as
\begin{eqnarray}
	q_1 < q_2 < q_3 < q_4  ,
\end{eqnarray}
corresponding to coordinates
\be\label{coo-outer}
s_1  =  \pi + {\rm arccos} \frac{x}{R}; \quad s_4  =  \pi - {\rm arccos} \frac{x}{R}
\ee
on the outer contour, and coordinates
\be
s_2  =  2\pi - {\rm arccos} \frac{x}{r}; \quad s_3  =   {\rm arccos} \frac{x}{r}
\ee
on the inner contour. Away from this region, i.e., for $r<|x|<R$, one finds a single Fermi sea with coordinates $s_1$, $s_4$ given in \eqref{coo-outer}. 

\begin{figure}[t!]
	\includegraphics[width=.975\columnwidth]{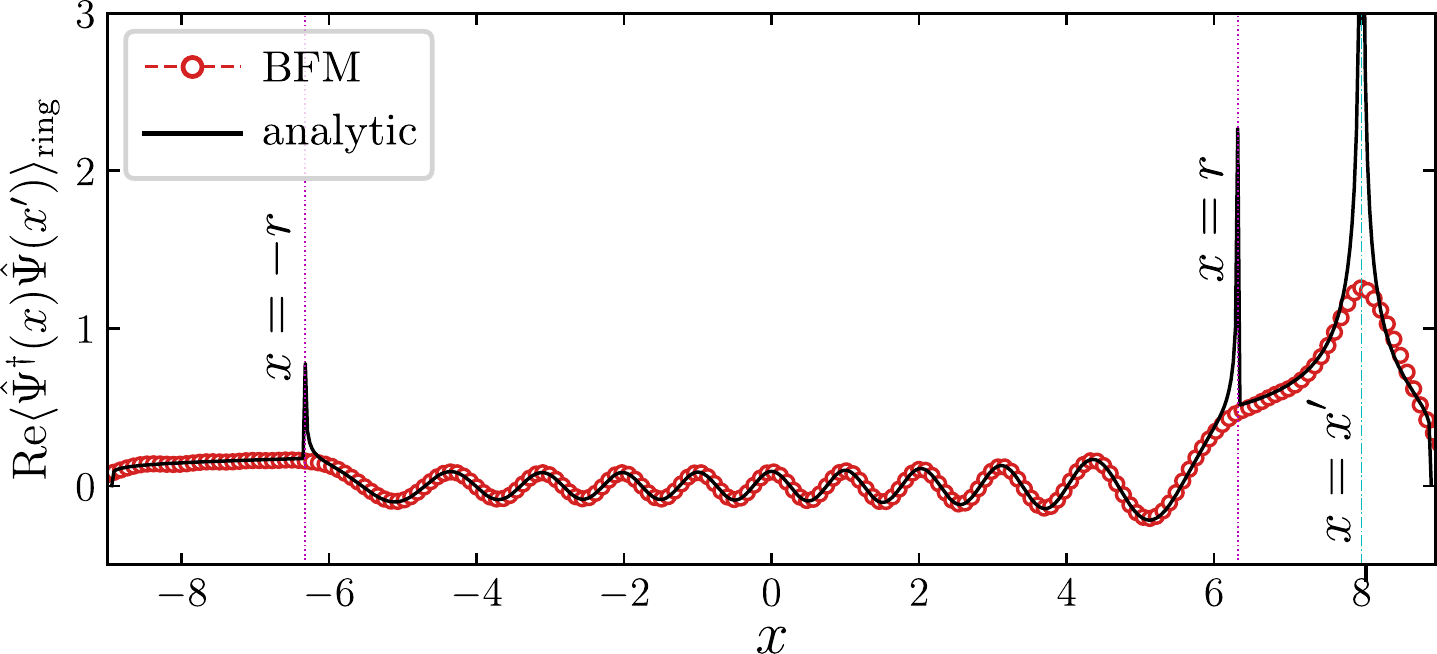}
\caption{1PDM density matrix for the ring state of Fig.~\ref{fig:ring} with $\omega=1$, $x'=8$ and $M=N=20$. The black full line shows the result obtained using Eq.~\eqref{eq:opdm-ring} and it is compared with BFM numerical results (red circles). We observe an overall excellent agreement of the two curves with divergences at $x=x'$ ({dash-dotted axes}) and for $x=\pm r$ ({dotted axes}).}\label{fig:ring-1pdm}
\end{figure}

The 1PDM is obtained using the formula in Eq.~\eqref{eq:mainresult}:
\be\label{eq:opdm-ring}
\langle \hat\Psi^\dagger(x)\hat\Psi(x')\rangle_\text{ring}= {\cal C}^\dagger(x) \cdot {\cal F}(x,x') \cdot {\cal C}(x). 
\ee
\begin{widetext}
For the $d_{Q_x}$-dimensional vector ${\cal C}^\dagger(x)$ we have the following results (hereafter $\hbar=\omega=1$):
\begin{itemize}
	\item[-] if $r < |x| < R$ (i.e., $Q_x=1$):
\begin{eqnarray}
	&&  \mathcal{C}^\dagger(x)  =  \left( \frac{G^2 (3/2)}{\sqrt{2\pi}} \right)\times 2^\frac{1}{4}    
\end{eqnarray}
	where the phase simplifies since $\Phi_o(s_1) + \Phi_o(s_4=2\pi-s_1) =0$.
	\item[-] if $ |x| < r$ (i.e., $Q_x=2$):
\be
 \mathcal{C}^\dagger(x)  =  \frac{\left( \frac{G^2 (3/2)}{\sqrt{\pi}} \right)^2}{\sqrt{2}}    
	 \left( \begin{array}{c}
		\frac{\exp\left(\frac{i}{2}[ -\Phi_1+\Phi_2+\Phi_3+\Phi_4]\right)}{(R^2-x^2)^{\frac{1}{4}}}  \\[3pt]
		\frac{\exp\left(\frac{i}{2}[ \Phi_1-\Phi_2+\Phi_3+\Phi_4]\right)}{(r^2-x^2)^{\frac{1}{4}}}   \\[3pt]
		\frac{\exp\left(\frac{i}{2}[ \Phi_1+\Phi_2-\Phi_3+\Phi_4]\right)}{(r^2-x^2)^{\frac{1}{4}}}   \\[3pt]
		\frac{\exp\left(\frac{i}{2}[ \Phi_1+\Phi_2+\Phi_3-\Phi_4]\right)}{(R^2-x^2)^{\frac{1}{4}}} 		
	\end{array} \right)  
\ee
where we used the shorthand $\Phi_o(s_{a=1,4})=\Phi_a$ and $\Phi_i(s_{b=2,3})=\Phi_b$.
\end{itemize}
For the $d_{Q_x}\times d_{Q_{x'}}$ matrix ${\cal F}(x,x')$, we find:
\begin{itemize} 
	\item[-] if $r < |x| < R$ and $r < |x'| < R$:
		\begin{equation}
			\mathcal{F}(x,x')= \frac{ R^{1/2} \left( 1- \frac{x^2}{R^2} \right)^{1/8}  \left( 1- \frac{x'^2}{R^2} \right)^{1/8} }{ \left| x-x'  \right|^{1/2} }
		\end{equation}

	\item[-] if $r < |x| < R$ and $|x'| < r $:
\begin{eqnarray}
	&&  \mathcal{F}(x,x')  =  \frac{\left( \frac{G^2 (3/2)}{\sqrt{\pi}} \right)^2}{\sqrt{2}}    
	 \left( \begin{array}{cccc}
		 0 &  \frac{R^{\frac{1}{2}}  \left( \frac{(1- \frac{x^2}{R^2} ) ( 1- \frac{x'^2}{R^2} ) }{ 1- \frac{x'^2}{r^2} } \right)^{1/8}  }{2^{\frac{1}{4}} |x-x'|^{\frac{1}{2}}}  &   \frac{R^{\frac{1}{2}}  \left( \frac{(1- \frac{x^2}{R^2} ) ( 1- \frac{x'^2}{R^2} ) }{ 1- \frac{x'^2}{r^2} } \right)^{1/8}  }{2^{\frac{1}{4}} |x-x'|^{\frac{1}{2}}}   & 0 
	\end{array} \right)  
\end{eqnarray}
	\item[-] if $ |x| < r$ and $r< |x'| < R $:
\begin{eqnarray}
	&&  \mathcal{F}(x,x')  =  \frac{\left( \frac{G^2 (3/2)}{\sqrt{\pi}} \right)^2}{\sqrt{2}}    
	 \left( \begin{array}{c}
		 0 \\  \frac{R^{\frac{1}{2}}  \left( \frac{(1- \frac{x^2}{R^2} ) ( 1- \frac{x'^2}{R^2} ) }{ 1- \frac{x'^2}{r^2} } \right)^{1/8}  }{2^{\frac{1}{4}} |x-x'|^{\frac{1}{2}}}  \\   \frac{R^{\frac{1}{2}}  \left( \frac{(1- \frac{x^2}{R^2} ) ( 1- \frac{x'^2}{R^2} ) }{ 1- \frac{x'^2}{r^2} } \right)^{1/8}  }{2^{\frac{1}{4}} |x-x'|^{\frac{1}{2}}}   \\ 0 
	\end{array} \right)  
\end{eqnarray}

	\item[-] if $ r<|x| < R$ and $r< |x'| < R $: 
\be\label{F-cumbersome}
	\mathcal{F}(x,x') ={\cal K}(x)
	\times \left( \begin{array}{cccc} 
		\frac{r^{\frac{1}{2}} }{ \sqrt{2}   |x-x'|^{\frac{1}{2} } }  \times \dots & 0 & 0 & \frac{r^{\frac{1}{2}} }{ \sqrt{2}   |x-x'|^{\frac{1}{2} } }  \times \dots \\
		0 &  \frac{R^{\frac{1}{2}} }{ \sqrt{2}   |x-x'|^{\frac{1}{2} } } \times \dots  &  \frac{R^{\frac{1}{2}} }{ \sqrt{2}   |x-x'|^{\frac{1}{2} } }   \times \dots & 0   \\
		0 &  \frac{R^{\frac{1}{2}} }{ \sqrt{2}   |x-x'|^{\frac{1}{2} } } \times \dots  &  \frac{R^{\frac{1}{2}} }{ \sqrt{2}   |x-x'|^{\frac{1}{2} } }  \times \dots & 0   \\
		\frac{r^{\frac{1}{2}} }{ \sqrt{2}   |x-x'|^{\frac{1}{2} } } \times \dots  & 0 & 0 & \frac{r^{\frac{1}{2}} }{ \sqrt{2}   |x-x'|^{\frac{1}{2} } }  \times \dots
	\end{array} \right) 
\times {\cal K}(x')
\ee
\end{itemize}
where
\be
{\cal K}(x)= \text{diag}\left( \begin{array}{cccc} 
		\left( \frac{1-x^2/r^2}{1-x^2/R^4} \right)^{\frac{1}{8}} \\
		 \left( \frac{1-x^2/R^2}{1-x^2/r^4} \right)^{\frac{1}{8}} \\
		 \left( \frac{1-x^2/R^2}{1-x^2/r^4} \right)^{\frac{1}{8}} \\
		 \left( \frac{1-x^2/r^2}{1-x^2/R^4} \right)^{\frac{1}{8}}
	\end{array} \right) 
\ee
and we omitted the full expression of the non-vanishing elements in \eqref{F-cumbersome} for a better exposition. In Fig.~\ref{fig:ring-1pdm}, we show the result for the 1PDM of the ring state with $M=20$ and $N=20$, compared to BFM numerical calculations performed with the method of Ref.~\cite{pezer2007momentum,atas2017exact}.\\
\end{widetext}
 Importantly, we observe that the matrix ${\cal F}(x,x')\propto |x-x'|^{-1/2}$ and it diverges in the limit of coincident points $x\to x'$, as expected within a field theory description. Moreover, additional power-law divergences of the 1PDM arise when one of the two points $|x|,|x'|=r$ i.e., when we pass from two to four Fermi points and viceversa. These divergences are related to the limit of coincident momenta $q_a\to q_b$  in a split Fermi sea at given position $x$ (and consequently $s_a\to s_b$) and affect both the non-universal amplitude of Eq.~\eqref{coef-FS} (which is notoriously ill-defined in the presence of non-distinct momenta) and the propagator of Eq.~\eqref{eq:propagator} . Both these types of divergences affecting the 1PDM can be regularized as explained in Appendix~\ref{sec:regularization}.
\vspace{0.3cm}
\begin{figure}[t!]
\includegraphics[width=.975\columnwidth]{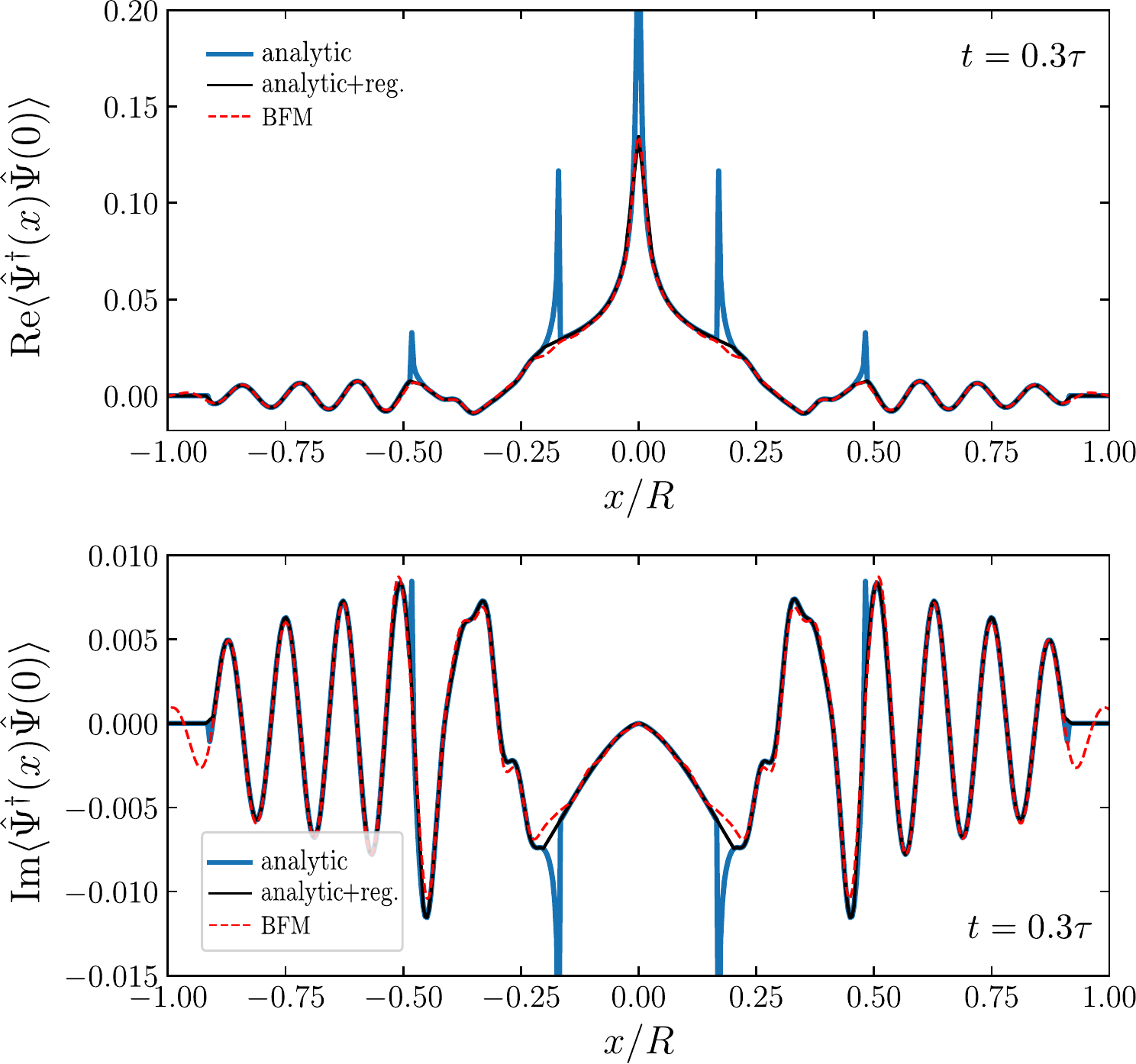}
\caption{Example of regularization of the 1PDM of Fig.~{\color{red}2} of the main text for $x'=0$ and at time $t=0.3\tau$ ({\it thick full line}). As one can see, by employing a simple linear interpolation scheme for the removal of the divergences, one finds already a very good agreement of the regularized 1PDM ({\it thin full line}) with the time-dependent BFM numerical data ({\it dashed line}).}\label{fig:qghd-reg}
\end{figure}

\section{Regularization of the 1PDM}\label{sec:regularization}
We finally discuss the regularization procedure for the divergences appearing in the 1PDM. Although all these divergences have a similar origin, we find it convenient to start from the divergence arising when $x\to x'$ and later move to the regularization of the secondary peaks of the 1PDM. As already commented in the main text, this divergence $g_1 (x, x') \equiv \langle \hat{\Psi}^\dagger(x)   \hat{\Psi}(x')  \rangle 
 \propto |x-x'|^{-1/2}$ characterizes the asymptotic behavior of the 1PDM already in homogeneous systems at equilibrium, which is indeed expected to break down at microscopic scales $|x-x'|\ll \rho^{-1}$. Nevertheless, short-distance expansions for the 1PDM have been systematically worked out for the Tonks-Girardeau gas exploiting  Fisher-Hartwig conjecture (see Ref.~\cite{jimbo1980density,forrester2003finite}). For instance, the first terms of this expansion read as
\be\label{eq:Jimbo}\begin{split}
g_1(r\equiv{2\pi x}/{L},0)&=\rho\Big\{1-\frac{(N^2-1)}{24}r^2+ \frac{N(N^2-1)}{72\pi}|r|^3 \\
&+ \frac{(3N^4-10 N^2 +7)}{5760} r^4 + O(|r|^5)\Big\}.
\end{split}\ee
Since in the limit $x\to x'$, our assumptions are compatible with a locally homogeneous fluid, one can then easily remove the divergence at $x\simeq x'$ by employing the expansion in Eq.~\eqref{eq:Jimbo} around the region $|x-x'|\ll \rho(x)^{-1}$. In practice, we experienced that even retaining only the few lowest terms in the expansion is enough to obtain a very good matching with the exact numerical data, see Fig.~\ref{fig:qghd-reg}.\\

Next, secondary peaks arise at turning points $s^*$ on the Fermi contour (i.e., when the number $Q$ of Fermi seas as function of real space position $x$ undergoes a discontinuity). Although these divergences manifest in the underlying field theory description in a similar fashion of that at $x=x'$, we find their regularization through a short-distance expansion similar to that in Eq.~\eqref{eq:Jimbo} a non-trivial calculation. Nevertheless, we observe that by employing a simple linear interpolation scheme for these divergences, namely, by linearly interpolating the values of $g_1(x,x')$ away from the divergence at $|x-x_t(s^*)|\simeq\delta$, with $\delta$ a constant $\sim O(1)$, we are already able to regularize the asymptotic result for the 1PDM in Eq.~\eqref{eq:mainresult} of the main text with good accuracy, see Fig.~\ref{fig:qghd-reg}. Indeed, this is also confirmed by the good agreement that we obtained for the MD of Fig.~\ref{fig:MD} of the main text, where the large momentum tails display only small deviations from the numerical data. These deviations can be minimized by improving our regularization scheme for the secondary peaks or by combining our asymptotic approach with local density approximation (see Ref.~\cite{caux2019hydrodynamics}), which is expected to become exact at large momentum. We plan to investigate these aspects in future publications.